\documentclass{ws-procs961x669}

\usepackage{xfrac}
\usepackage{amsmath}
\usepackage{braket}

\newcommand{\la}{\langle}
\newcommand{\ra}{\rangle}

\newcommand{\be}{\begin{equation}}
\newcommand{\ee}{\end{equation}}
\newcommand{\bea}{\begin{eqnarray}}
\newcommand{\eea}{\end{eqnarray}}
\newcommand{\bes}{\begin{subequations}}
\newcommand{\ees}{\end{subequations}}

\begin{document}

\title{Validity of the Semiclassical Approximation in 1+1 Electrodynamics: Numerical Solutions to the Linear Response Equation}

\author{Ian M. Newsome}

\address{Department of Physics, Wake Forest University, \\
Winston-Salem, NC 27109, USA \\
Email: newsim18@wfu.edu}

\begin{abstract}
From previous work \cite{validityEM}, the semiclassical backreaction equation in 1+1 dimensions was solved and a criterion was implemented to assess the validity of the semiclassical approximation in this case. The criterion involves the behavior of solutions to the linear response equation which describes perturbations about solutions to the semiclassical backreaction equation. The linear response equation involves a time integral over a two-point correlation function for the current induced by the quantum field and it is expected that significant growth in this two-point function (and therefore in quantum fluctuations) will result in significant growth in solutions to the linear response equation. It was conjectured for early times that the difference of two nearby solutions to the semiclassical backreaction equation, with similar initial conditions, can act as an approximate solution to the linear response equation. A comparative analysis between the approximate and numerical solutions to the linear response equation, for the critical scale for particle production, will be presented for the case of a massive, quantized spin \sfrac{1}{2} field in order to determine how robust the approximation method is for representing its solutions.
\end{abstract}


\section{Introduction}
The semiclassical approximation has been utilized in a wide range of physical situations where a quantized field evolves in a classical background. The relevant scenario for the following discussion involves the decay of an electric field via the Schwinger effect \cite{schwinger}. The original calculation by Schwinger in 1951 involved a background field calculation in which an electric field $E$, that is constant in space and time, gives rise to a particle production rate due to vacuum decay. At leading one-loop order the particle production rate is proportional to a factor of $\textnormal{exp}\{-\sfrac{\pi m^2}{qE}\}$, from which one can define a critical scale for pair production that is $E_{\textnormal{crit}} \sim \sfrac{m^2}{q}$.

The time evolution of the net electromagnetic field is described by the semiclassical backreaction equation and has been solved for the case of a homogeneous electric field coupled to a massive scalar or a massive spin \sfrac{1}{2} field in 1+1 \cite{Kluger91,Kluger92,Kluger93} and 3+1 \cite{Kluger93, Tanji, stat-FT} dimensions. When particles are produced, they accelerate in reaction to the background electric field, creating an electric current. This current produces a counter electric field which initially damps the original background electric field, and at late times the net electric field will oscillate.

Given that the semiclassical electrodynamics model is an approximation to quantum electrodynamics, it must be asked to what extent is this an accurate approximation? Concerning the model presented here, the semiclassical backreaction equation involves an expectation value of the current constructed from quantum fields and for this to be a satisfactory approximation to what one would measure, it is necessary that quantum fluctuations be small.

To summarize the previous work which led to the development of the numerical solution to the linear response equation presented in these proceedings, the main goals in [1] had been to study the details of the particle production process when backreaction effects were taken into account and subsequently to estimate the significance of certain types of quantum fluctuations to investigate the validity of the semiclassical approximation for 1+1 electrodynamics. This included solving the necessary backreaction equations for scalar and spin \sfrac{1}{2} fields coupled to two different classical source profiles, one which led to an asymptotically constant electric field and the other being the Sauter pulse. To wit, a criterion was implemented to assess the validity of the semiclassical approximation which had been previously applied to the process of preheating in models of chaotic inflation \cite{validitypreheating}, with an earlier version applied to semiclassical gravity \cite{validitygravity}. This criterion involves the behavior of solutions to the linear response equation, which can be derived by perturbing the semiclassical backreaction equation about a background field solution.

In general, the linear response equation is a second order integro-differential equation which can be cumbersome to solve numerically. Therefore a method was developed \cite{validitypreheating} to approximate solutions to the linear response equation which involves the difference between solutions to the semiclassical backreaction equation with similar initial conditions. For early times, this difference is expected to act as an approximate solution to the linear response equation as long as the exact solution is relatively small.

Subsequently, an effort has been made to directly solve for the numerical solutions to the linear response equation for the case of an asymptotically constant background electric field, initially zero, coupled to a spin \sfrac{1}{2} field. For the purposes of this proceeding, the numerical solutions to the linear response equation are desired in order to determine how robust the method for approximating its solutions will be. Therefore, a comparative analysis between the numerical solutions to the linear response equation and its approximate solutions described previously will be presented in what follows, with special attention given to the critical scale for particle production.


The structure of this paper is as follows: in Sec.\,2 a brief review of the system is discussed including quantization of the spin \sfrac{1}{2} field, the semiclassical backreaction equation, and the renormalization procedure used. Sec.\,3 contains a review of relevant material \cite{validityEM} including the criterion for the validity of the semiclassical approximation with the general and specific forms of the linear response equation presented for the case of a spin \sfrac{1}{2} field coupled to a background electric field. Also, included is a discussion of the method used to approximate solutions to the linear response equation. In Sec.\,4 an analysis of the numerical solutions to the linear response equation is given, with a comparison drawn between its approximate solutions, with an emphasis on the critical scale for significant particle production. Sec.\,5 includes a summary of results and discussion thereof.

\section{Quantization and Renormalization of a Spin \sfrac{1}{2} Field}
This section describes the model under consideration: a quantized spin \sfrac{1}{2} field which interacts with a background electric field generated by a prescribed classical source. Analysis is restriced to 1+1 Minkowski spacetime with the assumption that the background electric field is spatially homogeneous in a given reference frame, i.e. $E=E(t)$. Furthermore, the metric signature is chosen to be $(-,+)$ with $c=\hbar=1$.

The classical action describing a spin \sfrac{1}{2} field $\psi(t,x)$ coupled to a background electric field is given by
\be
    S = \int d^2x \left[ -\frac{1}{4}F_{\mu \nu}F^{\mu \nu} + A_\mu J^\mu_C + i \Bar{\psi} \gamma^{\mu} D_\mu \psi - m \Bar{\psi} \psi \right] \quad . \label{spin1/2action}
\ee
Here $F_{\mu \nu}=\partial_\mu A_\nu - \partial_{\nu}A_{\mu}$ is the electromagnetic field strength tensor, $m$ the mass of spin \sfrac{1}{2} field excitations, $D_{\mu}=\partial_{\mu}-i q A_{\mu}$ the gauge covariant derivative, $J^\mu_C$ a classical and conserved external source, and $\Bar{\psi}=\psi^{\dagger}\gamma^0$. The Dirac matrices $\gamma^{\mu}$ satisfy the following anticommutation relations $\{\gamma^\mu , \gamma^\nu\}=-2 \eta_{\mu \nu}$.

Variation of \eqref{spin1/2action} with respect to the vector potential yields the general form of Maxwell's equation
\be
    -\Box A^\mu + \partial^\mu \partial_\nu A^\nu = J^\mu_C + J^\mu_Q \quad . \label{Max}
\ee
The classical source and background electric field generated by this source is chosen to be
\be
    J_C = - \frac{q E_0}{(1+qt)^2} \quad , \quad E_C = -\int J_C \, dt = E_0 \left( \frac{qt}{1 + qt} \right) \quad , \label{jclass}
\ee
for $t\geq 0$ and $J_C=0$ for $t<0$. Note, the term $J_C$ in \eqref{jclass} is the spatial component of $J^\mu_C$. Since the classical current is initially zero, and gives rise to an electric field that is initially zero as well, there is no ambiguity in the choice of vacuum state. The source term $J^\mu_Q$ induced by the spin \sfrac{1}{2} field is given by
\be
    J^\mu_Q = q \Bar{\psi} \gamma^\mu \psi \quad .
\ee
Variation of \eqref{spin1/2action} with respect to the field $\Bar{\psi}$ yields the Dirac equation
\be
    \left(i\, \gamma^\mu D_\mu - m\right)\psi(t,x) = 0 \quad .
\ee
In what follows, the gauge choice
\be
    A^\mu = (0,A(t)) \quad , \label{gauge}
\ee
will be implemented. Expanding the spin \sfrac{1}{2} field $\psi(t,x)$ in terms of a complete set of modes yields
\be
    \psi(t,x) = \int_{-\infty}^{\infty} dk \, \left[ B_k u_k(t,x) + D_k^{\dagger} v_k(t,x) \right] \quad , \label{field}
\ee
with $B_k , B_k^{\dagger} , D_k , D_k^{\dagger}$ the usual creation and annihilation operators obeying the anticommutation relations $\{B_k , B_{k^{'}} \} = \{D_k , D^{\dagger}_{k^{'}} \} = \delta (k-k^{'})$. Utilizing a particular form for the modes \cite{FerrSalas}, two independent spinor solutions can be constructed as follows
\be
    u_k(t,x) = \frac{e^{i k x}}{\sqrt{2\pi}}
    \begin{pmatrix}
        h_k^{I}(t) \\ \\
        -h_k^{II}(t)
    \end{pmatrix}
    \quad , \quad v_k(t,x) = \frac{e^{-i k x}}{\sqrt{2\pi}}
    \begin{pmatrix}
        h_{-k}^{II\, *}(t) \\ \\
        h_{-k}^{I\, *}(t)
    \end{pmatrix} \quad .
\ee
Using the Weyl representation of the Dirac matrices $\gamma^\mu$
\be
    \gamma^0 = 
    \begin{pmatrix}
        0 & 1 \\
        1 & 0
    \end{pmatrix} \quad , \quad
    \gamma^1 =
    \begin{pmatrix}
        0 & 1 \\
        -1 & 0
    \end{pmatrix} \quad ,
\ee
the functions $h_k^{I}(t)$ and $h_k^{II}(t)$ satisfy the mode equations
\bes
    \be
        \Dot{h}_k^I - i (k - qA) h_k^I - i m h_k^{II} = 0 \quad , \label{mode1} \\
    \ee
    \be
        \Dot{h}_k^{II} + i (k - qA) h_k^{II} - i m h_k^{I} = 0 \quad . \label{mode2}
    \ee
\ees

The time evolution of this system is governed by the semiclassical backreaction equation. This can be obtained by replacing $J^\mu_Q$ in \eqref{Max} with its expectation value $\la J^\mu_Q \ra$ and then use the gauge choice \eqref{gauge} with \eqref{field} to yield
\be
    \frac{d^2}{dt^2}A(t) = -\frac{d}{dt}E(t) = J_C + \la J_Q \ra \quad . \label{SBE}
\ee
Due to the coupling between the classical background electric field and the quantized spin \sfrac{1}{2} field, charged spin \sfrac{1}{2} particles will be created. These particles will be accelerated by the background electric field, creating a current which generates a secondary electric field. In the semiclassical approximation, the current created from the accelerated spin \sfrac{1}{2} particles is given by $\la J_Q \ra$. The renormalized expression for $\la J_Q \ra$, evaluated in the vacuum state is \cite{validityEM}
\be
    \la J_{Q} \ra_{\textnormal{ren}} = \frac{q}{2\pi}\int_{-\infty}^{\infty} dk \, \left[ |h_{k}^{I}(t)|^{2}-|h_{k}^{II}(t)|^{2} + \frac{k}{\omega} - \frac{q \, m^2}{\omega^3}A(t) \right] \quad . \label{Jrenorm}
\ee
with $\omega^2 = k^2 + m^2$. Here adiabatic regularization was used to eliminate the ultraviolet divergences \cite{FerrSalas}.

\section{Validity Criterion for the Semiclassical Approximation}
This section gives a review of relevant material \cite{validityEM} required for the investigation of the solutions to the linear response equation.

Since the current term $\la J_Q \ra$, in part, characterizes the quantum particle production process, a natural way to determine the size of quantum fluctuations, compared to other relevant quantities, is to evaluate a two-point correlation function for the current. In general, there are a number of different correlation functions which could be used, but in order to avoid such problems as state-dependent divergences \cite{wu-ford}, incompatible results from various renormalization techniques \cite{phillips-hu}, or covariance issues \cite{validitygravity}, it is useful to proceed with a two-point correlation function which emerges naturally from the semiclassical theory itself, namely $\la [J_Q(t,x) , J_Q(t^{'},x^{'})] \ra$. This two-point function measures the extent to which the value of the current $\la J_Q \ra$ at two separate spacetime points commutes, thereby having the interpretation of characterizing the degree to which quantum fluctuations are introduced into the system.

Perturbing the semiclassical backreaction equation yields the linear response equation which contains this two-point correlation function and describes the time evolution of perturbations about a given semiclassical solution. A criterion for the validity of the semiclassical approximation was originally developed for semiclassical gravity \cite{validitygravity} and modified for preheating in chaotic inflation \cite{validitypreheating}. This criterion was applied to semiclassical electrodynamics \cite{validityEM} and states: the semiclassical approximation will break down if any linearized gauge invariant quantity constructed from solutions to the linear response equation with finite non-singular initial data grows rapidly for some period of time. It is important to note this is a necessary, but not sufficient condition for the validity of the semiclassical approximation.

\subsection{The Linear Response Equation}
The linear response equation for semiclassical electrodynamics is found by perturbing the semiclassical backreaction equation about a given solution. From \eqref{SBE} this becomes
\be
    \frac{d^2}{dt^2}\delta A(t) = -\frac{d}{dt}\delta E(t) = \delta J_C + \delta \la J_Q \ra \quad . \label{LRE}
\ee
More specifically, the type of perturbation being performed is one that changes the classical current by altering the value of the classical background electric field amplitude $E_0$ in \eqref{jclass}. Thus the term $\delta J_C$ is expressed as
\be
    \delta J_C = - \frac{q}{(1+qt)^2} \delta E_0 \quad . \label{deltaJclass}
\ee

In conjunction with the validity criterion stated previously, it is useful to break up the solutions to the semiclassical backreaction equation into the purely classical and quantum pieces
\bes
    \be
        E_C = -\int_{t_0}^{t} J_C(t_1) \, dt_1 \quad , \label{Ec}
    \ee
    \be
        E_Q = E - E_C \quad . \label{Eq}
    \ee
\ees
Solutions to the linear response equation can be partitioned in the same way which allows for the statement of the validity criterion to be modified to state that if $\delta E_Q$ grows rapidly during some time interval, then the semiclassical approximation is not valid.

For the case of a spin \sfrac{1}{2} field, the renormalized $\delta \la J_Q \ra$ term present in \eqref{LRE} can be expressed as \cite{validityEM}
\be
    \delta \la J_Q \ra _{\textnormal{ren}} = -\frac{q^2}{\pi} \delta A(t) + i \int_{-\infty}^{\infty} dx^{'} \int_{-\infty}^{t} dt^{'} \, \la [J_Q(t,x) , J_Q(t^{'},x^{'})] \ra \, \delta A(t^{'}) \quad ,
\ee
with the relationship
\be
    \int_{-\infty}^{\infty} dx^{'} \, \la [J_Q(t,x) , J_Q(t^{'},x^{'})] \ra = \frac{4 i q^2}{\pi} \int_{-\infty}^{\infty} dk \, \textnormal{Im}\left\{h_k^I(t)h_k^{II}(t)h_k^{I*}(t^{'})h_k^{II*}(t^{'})\right\} \quad . \label{twopoint}
\ee
Thus \eqref{LRE} takes the specific form
\bea
    \frac{d^2}{dt^2}\delta A(t) &=& - \frac{q}{(1+qt)^2}\delta E_0 - \frac{q^2}{\pi} \delta A(t) \nonumber \\
    && - \frac{4 q^2}{\pi} \int_{-\infty}^{\infty} dk \int_{-\infty}^{t} dt^{'} \,  \textnormal{Im}\left\{h_k^I(t)h_k^{II}(t)h_k^{I*}(t^{'})h_k^{II*}(t^{'})\right\} \delta A(t^{'}) \quad . \nonumber \\ \label{fullLRE}
\eea

\subsection{Approximate Solutions to the Linear Response Equation}
A technique that was developed to approximate solutions to the linear response equation for the case of homogeneous perturbations \cite{validityEM} involves solving the semiclassical backreaction equation for two solutions whose initial conditions are similar. It is expected that the difference between these two solutions $\Delta E$ is an approximate solution to the linear response equation $\delta E$, for early times, as long as $\Delta E \approx \delta E$. If the difference does grow large, then the corresponding solution to the linear response equation should grow significantly as well. This would violate the criterion discussed above and thus signal a breakdown in the semiclassical approximation.

A way to measure the relative growth of two solutions $E_1$ and $E_2$ to the semiclassical backreaction equation is with the modified relative difference expression \cite{validityEM}
\be
    R = \frac{|\Delta E|}{|E_1| + |E_2|} \quad , \quad \Delta E = E_2 - E_1 \quad .
\ee
From the semiclassical backreaction equation, it is clear the difference $\Delta E$ will be a solution of the following equation
\be
    -\frac{d}{dt}\Delta E = \Delta J_C + \Delta \la J_Q \ra \quad ,
\ee
with the definitions
\be
    \Delta J_C = J_{C,2} - J_{C,1} \quad , \quad \Delta \la J_Q \ra = \la J_{Q,2} \ra - \la J_{Q,1} \ra \quad .
\ee
For $\Delta E$ to be an approximate solution to the linear response equation, i.e. $\Delta E \approx \delta E$, it is clear that $\Delta \la J_Q \ra \approx \delta \la J_Q \ra$ must hold, since one can set $\Delta J_C = \delta J_C$ for all times.

A natural way to measure the growth of the approximate quantum contribution to the finite difference equation $\Delta E_Q$ is with a relative difference $R_Q$ given as
\be
    R_Q = \frac{|\Delta E_Q|}{|E_{Q,1}| + |E_{Q,2}|} \quad , \quad \Delta E_Q = E_{Q,2} - E_{Q,1} \quad .
\ee
This acts as the gauge invariant quantity constructed from solutions to the linear response equation mentioned in the formal statement for the validity criterion. This difference can be compared to the relative difference $R_C$ between corresponding classical solutions, which does not vary in time. Therefore if $R_Q$ grows rapidly for some period of time, this signals a breakdown in the semiclassical approximation.

\begin{figure}[h]
    \centering
    \includegraphics[scale=0.75]{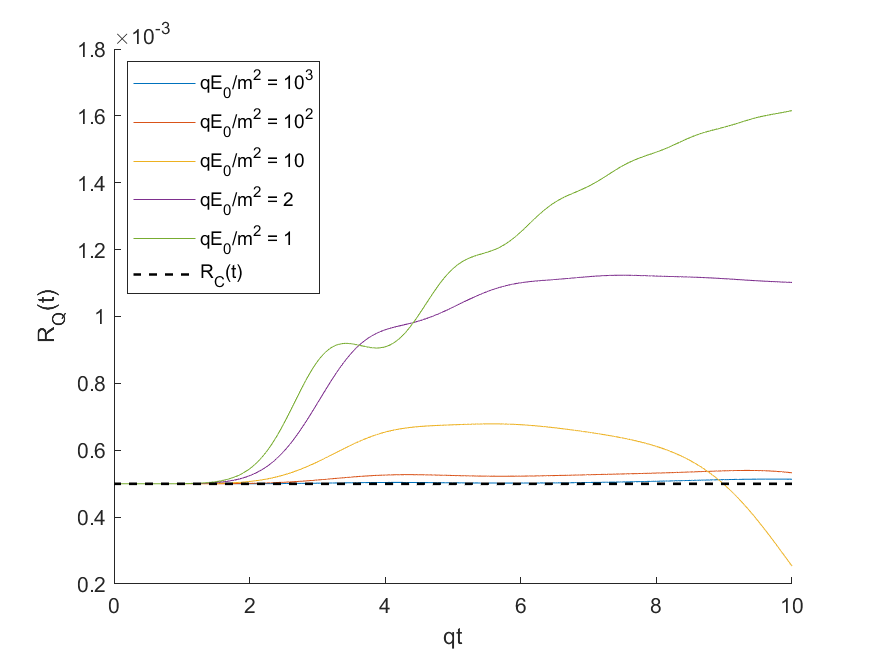} 
    \caption{Results obtained for the quantity $R_Q$. The values $E_{0,1}/q=1$ and $E_{0,2}/q=1+10^{-3}$ have been chosen for the solutions which comprise $\Delta E$. The values of the characteristic quantity $qE_0/m^2$ for each case are shown along with the type of curve for that solution in the legend. Here $R_C(t)$ denotes the relative difference excluding quantum effects.}
    \label{Rqplot}
\end{figure}

In Fig. \ref{Rqplot}, some results \cite{validityEM} are given for the quantity $R_Q$ for a variety of different particle masses\footnote{This is an aggregate plot built from the results and data presented in [1].}. The most significant effect on the behavior of $R_Q$ is the size of the characteristic dimensionless quantity $\sfrac{qE_0}{m^2}$. It is therefore useful to distinguish between three different regimes: (i) $\sfrac{qE_0}{m^2}\gg 1$ in which the mass is relatively small compared to the electric field, resulting a large amount of particle production, (ii) $\sfrac{qE_0}{m^2}\sim 1$ in which the mass is of the same order as the electric field, resulting in less but still significant particle production, and (iii) $\sfrac{qE_0}{m^2}\ll 1$ in which the mass is relatively large compared to the electric field, resulting in very little particle production. The critical case for particle production is defined as $E \sim E_{\textnormal{crit}}=m^2/q$ and therefore the most relevant mass value is $m^2/q^2=1$. 

As seen in Fig.~\ref{Rqplot}, this particular mass value corresponds to the most rapid growth of $R_Q$ based on the cases considered here, and this implies the most severe breakdown of the semiclassical approximation for this case. However, one can see that $R_Q$ has the least amount of growth in the small mass regime and steadily increases its early time growth as the mass is increased up to the critical case. The conclusion drawn in [1], in the context of the validity criterion put forward, is that the semiclassical approximation is most greatly violated when $\sfrac{qE_0}{m^2}\sim 1$ and becomes more accurate as the particle mass is decreased (or increased since quantum fluctuations are expected to diminish due to lack of particle production in the high mass regime).

\section{Numerical Solutions to the Linear Response Equation}
In this section, an analysis of the numerical solution to the linear response equation for the critical case in which the characteristic quantity $\sfrac{qE_0}{m^2}=1$ is presented. A comparison is then drawn between its approximate and numerical solutions in the context of the validity criterion stated previously.

To arrive at numerical solutions to the linear response equation, one can partition the problem into two parts. The first part involves solving the semiclassical backreaction equation, and the second involves using the output from the semiclassical backreaction equation to supply the necessary data for solving the linear response equation. Generating the solutions to the semiclassical backreaction equation involves simultaneously solving a set of three coupled differential equations, namely \eqref{mode1}, \eqref{mode2}, and \eqref{SBE} with \eqref{Jrenorm}. The solution output will be the vector potential $A(t)$, and by extension the electric field $E(t)$, and the time dependent spin \sfrac{1}{2} field modes $h^{I,II}_k(t)$. The quantities needed for solving the linear response equation \eqref{fullLRE} are the modes $h^{I,II}_k(t)$, which the two-point function in \eqref{twopoint} depends on.

\subsection{Numerical Details}
The $\ket{in}$ state of the time dependent modes $h^{I,II}_k (t)$, prior to the classical source being turned on, is defined as positive frequency plane wave solutions to the mode equation. When the source term is turned on at $qt=0$ and subsequently $A(t)\neq 0$, the mode solutions are altered by the presence of the background field, specifically $k\rightarrow k-qA(t)$. This leads to particle production which is measured, in part, by the quantity $\la J_Q \ra$. Since the two-point function $\la [ J_Q(t,x) , J_Q(t^{'},x^{'}) ] \ra$ depends on these time dependent modes in a non-trivial way, the modification to the mode solutions will induce a non-zero contribution to the two-point function.

It is useful to separate the time dependent modes $h^{I,II}_k(t)$ into their real and imaginary parts
\bes
    \be
        h^I_k = \textnormal{Re}\{h^I_k\} + i \, \textnormal{Im}\{h^I_k\} \quad , \label{tmode1}
    \ee
    \be
        h^{II}_k = \textnormal{Re}\{h^{II}_k\} + i \, \textnormal{Im}\{h^{II}_k\} \quad . \label{tmode2}
    \ee
\ees
From this, the two mode equations defined in \eqref{mode1} and \eqref{mode2} now become four mode equations, given as
\bes
    \be
        \textnormal{Re}\{\dot{h}^{I}_k\} + (k-qA) \, \textnormal{Im}\{h^{I}_k\} + m \, \textnormal{Im}\{h^{II}_k\} = 0 \quad ,
    \ee
    \be
        \textnormal{Im}\{\dot{h}^{I}_k\} - (k-qA) \, \textnormal{Re}\{h^{I}_k\} - m \, \textnormal{Re}\{h^{II}_k\} = 0 \quad ,
    \ee
    \be
        \textnormal{Re}\{\dot{h}^{II}_k\} - (k-qA) \, \textnormal{Im}\{h^{II}_k\} + m \, \textnormal{Im}\{h^{I}_k\} = 0 \quad ,
    \ee
    \be
        \textnormal{Im}\{\dot{h}^{II}_k\} + (k-qA) \, \textnormal{Re}\{h^{II}_k\} - m \, \textnormal{Re}\{h^{I}_k\} = 0 \quad .
    \ee
\ees
Furthermore, the quantity $\la J_Q \ra$ in \eqref{Jrenorm} can now be expressed in the following way
\be
    \langle J_Q \rangle_{\textnormal{ren}} = - \frac{q^2}{\pi}A(t) + \frac{q}{2\pi}\int_{-\infty}^{\infty} dk \bigg[ \textnormal{Re}^2\{h^{I}_k\} + \textnormal{Im}^2\{h^{I}_k\} - \textnormal{Re}^2\{h^{II}_k\} - \textnormal{Im}^2\{h^{II}_k\} + \frac{k}{\omega} \bigg] \, .
\ee
It is straightforward to now implement a numerical routine which will solve Eqs. (10) and \eqref{SBE}.

From \eqref{twopoint} it is clear that $\la [J_Q(t,x),J_Q(t^{'},x^{'})] \ra$ depends on the modes $h^{I,II}_k(t)$. However, the time integral present in \eqref{fullLRE} depends on both the current time $t$ and the integration variable $t^{'}$. In order to make this integral suitable for numerical evaluation, it is convenient to utilize \eqref{tmode1} and \eqref{tmode2} which allows it to be expressed as
\bea
    && \int_{-\infty}^{t}dt^{'} \, \textnormal{Im}\big\{ h_k^I (t)h_k^{II} (t) h_k^{I*} (t^{'}) h_k^{II*} (t^{'}) \big\} \delta A(t^{'}) \nonumber \\
    && \quad = \bigg( \textnormal{Im}\big\{h_k^I (t)\big\}\textnormal{Im}\big\{h_k^{II} (t)\big\} - \textnormal{Re}\big\{h_k^I (t)\big\}\textnormal{Re}\big\{h_k^{II} (t)\big\} \bigg) \nonumber \\
    && \qquad \, \, \, \times \int_{-\infty}^{t}dt^{'} \, \bigg( \textnormal{Re}\big\{h_k^{I*} (t^{'})\big\}\textnormal{Im}\big\{h_k^{II*} (t^{'})\big\} + \textnormal{Im}\big\{h_k^{I*} (t^{'})\big\}\textnormal{Re}\big\{h_k^{II*} (t^{'})\big\} \bigg) \delta A(t^{'}) \nonumber \\
    && \qquad \, \, \, + \bigg( \textnormal{Re}\big\{h_k^I (t)\big\}\textnormal{Im}\big\{h_k^{II} (t)\big\} + \textnormal{Im}\big\{h_k^I (t)\big\}\textnormal{Re}\big\{h_k^{II} (t)\big\} \bigg) \nonumber \\
    && \qquad \, \, \, \times \int_{-\infty}^{t}dt^{'} \, \bigg( \textnormal{Re}\big\{h_k^{I*} (t^{'})\big\}\textnormal{Re}\big\{h_k^{II*} (t^{'})\big\} - \textnormal{Im}\big\{h_k^{I*} (t^{'})\big\}\textnormal{Im}\big\{h_k^{II*} (t^{'})\big\} \bigg) \delta A(t^{'}) \, \, \, . \nonumber \\
\eea
Here the inter-dependence of $t$ and $t^{'}$, seen in the product $h_k^I (t)h_k^{II} (t) h_k^{I*} (t^{'}) h_k^{II*} (t^{'})$, has been separated out. This allows for the time integral to be computed which then acts as the integrand for the $k$-integral in \eqref{fullLRE}, which can be computed as well. From this the numerical solutions $\delta A(t)$ and $\delta E(t)$ to the linear response equation can be found.

\subsection{Results and Discussion}
\begin{figure}[h]
    \centering
    \includegraphics[scale=0.29]{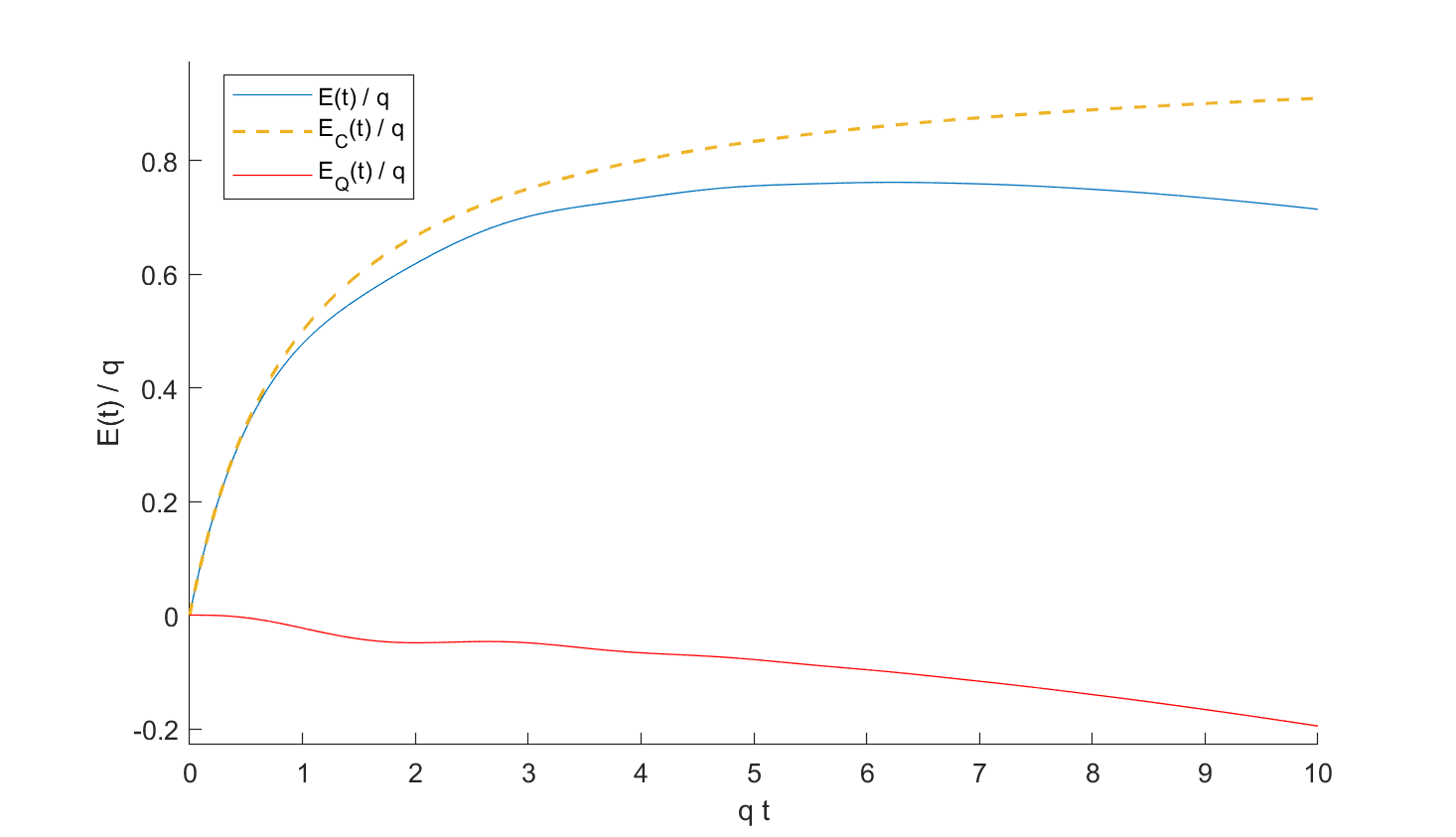}
    \caption{A plot of the electric field including backreaction. The net electric field $E(t)$ is given by the blue curve, the electric field $E_C(t)$ with no quantum effects present is given by the yellow dashed curve, and the electric field $E_Q(t)$ generated by only quantum effects is given by the red curve.}
    \label{Efield}
\end{figure}

In what follows, numerical results are presented for the critical case of $\sfrac{q E_0}{m^2}=1$. In Fig. \ref{Efield}, the net electric field $E(t)$ including backreaction is plotted \cite{validityEM} with both the electric field contributions $E_C$ when quantum effects are absent and $E_Q$ generated by only quantum effects, seen in \eqref{Eq} and \eqref{Ec}. One can see that at early times the net electric field begins to dampen compared to its classical counterpart $E_C$. This is due to the backreaction of the produced particles, resulting from the coupling of the spin $\sfrac{1}{2}$ field to the classical background source, causing an increase in the field $E_Q$.

\begin{figure}[h]
    \centering
    \includegraphics[scale=0.3]{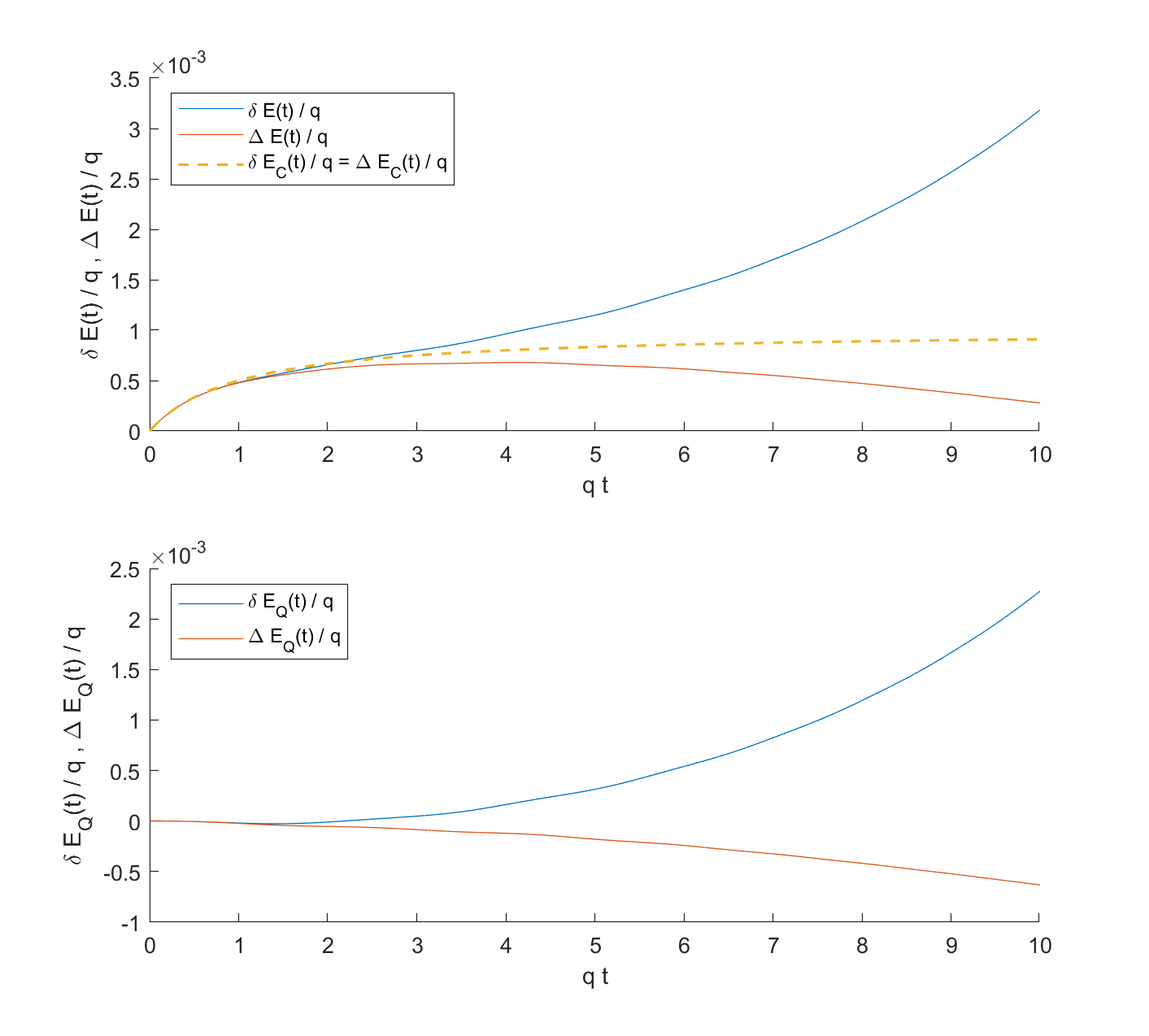}
    \caption{The top plot shows solutions to the linear response equation, with red and blue curves representing the approximate and numerical solutions $\Delta E$ and $\delta E$, respectively, with the yellow dashed curve being the classical contribution to the solution. The bottom plot isolates the quantum contribution to the linear response equation, for both the approximate and numerical solutions $\Delta E_Q$ and $\delta E_Q$. All plots reflect the critical case of $\sfrac{qE_0}{m^2}=1$ and were done with $\delta E_0$ and $\Delta E$ of order $10^{-3}$ for early times.}
    \label{deltaE_DeltaE}
\end{figure}

Numerical solutions to the linear response equation are plotted in Fig. \ref{deltaE_DeltaE}. The top plot shows the numerical solution $\delta E$ to the linear response equation, its approximate solution given by the finite difference between two nearby semiclassical backreaction equation solutions $\Delta E$, and the classical pieces $\delta E_C = \Delta E_C$. The bottom plot shows the purely quantum contributions to both the approximate $\Delta E_Q$ and numerical $\delta E_Q$ solutions to the linear response equation. From \eqref{deltaJclass} the initial value $\delta E_0$ was chosen in such a way as to equal the finite difference $\Delta E$ at early times, with this value being of order $10^{-3}$. 

For early times, there appears to be agreement between $\delta E$ and $\Delta E$, as well as $\Delta E_Q$ and $\delta E_Q$ since the classical contribution dominates due to a lack particle production. However, near the time $qt=1$ significant deviation between the two begins to occur, as could be measured by $(\delta E - \Delta E)/\delta E$. This deviation and subsequent late time growth in $\delta E$ is driven by the dependence of $\la [ J_Q(t,x) , J_Q(t^{'},x^{'}) ] \ra$ on the modes $h^{I,II}_k(t)$, whose positive frequency plane wave solutions are being altered due to the presence of the classical source term $A(t)$ having been switched on and leading to particle production. Since the critical case $\sfrac{q E_0}{m^2}=1$ considered is a threshold case for significant particle production, above which even more particle production will occur, the growth of $\delta E$ due to $\la [ J_Q(t,x) , J_Q(t^{'},x^{'}) ] \ra$ is expected.


The time $qt=1$ is also when one begins to notice substantial deviation between the net and classical electric fields, seen in Fig. \ref{Efield}, due to the presence of a nontrivial $\la J_Q \ra$ factor in \eqref{SBE} which will grow as more particles are produced. Furthermore, this time is also when one sees substantial growth of $R_Q$ in Fig. \ref{Rqplot} for the critical case. It therefore appears that for early times the $\Delta E$ is a good approximation to the linear response equation solution $\delta E$, and for late times $\delta E \neq \Delta E$. Comparing to the corresponding $R_Q$ behavior, there is a breakdown in the semiclassical approximation based on the validity criterion implemented here. This was the conclusion that was arrived at in [9], and it appears the behavior of the solutions $\delta E$ as compared with the $\Delta E$ validates this conclusion.

\section{Conclusions}
In previous work \cite{validityEM}, a criterion for the validity of the semiclassical approximation was applied to the case of 1+1 semiclassical electrodynamics which involves the behavior of solutions to the linear response equation. The linear response equation depends, in part, on a two-point correlation function $\la[J_Q(t,x),J_Q(t^{'},x^{'})\ra$, which characterizes the quantum fluctuations introduced through the current $\la J_Q \ra$. Numerical solutions to the linear response equation have been obtained for 1+1 semiclassical electrodynamics using a model of the Schwinger effect in which particle production occurs in the presence of a strong, spatially homogeneous, electric field. The particle production, in the context of the system considered here, is a consequence of the coupling between a quantized spin \sfrac{1}{2} field and a classical, asymptotically constant, background electric field generated by an external source. Since the linear response equation depends on $\la [ J_Q(t,x) , J_Q(t^{'},x^{'}) ] \ra$, if there is rapid growth in this correlation function for some period of time, then quantum fluctuations must be significant, and will drive the growth of solutions $\delta E$ to the linear response equation, resulting in a breakdown of the semiclassical approximation.

A previously developed technique to approximate solutions to the linear response equation involves the difference $\Delta E$ between two solutions to the semiclassical backreaction equation with similar initial conditions. The behavior of a quantity $R_Q$ built from this difference of semiclassical backreaction equation solutions gives a measure for the validity of the semiclassical approximation. When $R_Q$ grows large over some time, the corresponding linear response equation solution is expected to do so as well.

An analysis comparing the numerical $\delta E$ and approximate $\Delta E$ solutions to the linear response equation has been conducted. At the critical scale, where $\sfrac{qE_0}{m^2}\sim 1$, it was found that at early times the quantity $\Delta E$ approximated the linear response equation solutions $\delta E$ quite well. It was only after enough time had passed and significant particle production occurred that the numerical and approximate solutions deviated from one another. Therefore the claim that $\Delta E$ adequately approximates solutions to the linear response equation for early times is substantiated in this case. This critical scale was also the case for which the quantity $R_Q$ had its largest growth \cite{validityEM} and therefore was the case considered for the purposes of this proceeding. The significant growth in both $\delta E$ and $R_Q$ signals a breakdown of the semiclassical approximation based on the validity criterion utilized here.

Regarding future work, the numerical solutions to the linear response equation will be used to further investigate the nature of how quantum fluctuations affect these solutions. More specifically, since the linear response equation involves several factors which dictate the behavior of its solutions, one of which is a term characterizing quantum fluctuations introduced through the particle production process, having the numerical solution will allow one to isolate and investigate its respective contributions in detail.

\section*{Acknowledgements}
I. M. N. would like thank Silvia Pla for sharing semiclassical backreaction numerical code and data as well as helpful discussions, Eric Grotzke for helpful discussions regarding linear response equation numerical details, Paul R. Anderson for guidance regarding the creation of this manuscript as well as helpful discussions, and Jose Navarro-Salas and Kaitlin Hill for helpful discussions. All numerical work was performed using MATLAB software with special thanks to the Distributed Environment for Academic Computing (DEAC) at Wake Forest University for providing HPC resources which contributed to the research presented in this proceeding.

\end{document}